\documentclass[conference]{IEEEtran}
\usepackage{amsmath}
\usepackage{amsthm}
\usepackage{amsfonts}
\usepackage{amssymb}
\usepackage{graphicx}
\usepackage{color}
\usepackage{bm}
\usepackage[capitalize]{cleveref}
\usepackage{cite}

\usepackage[left=0.68in,right=0.68in,top=0.7in,bottom=1in]{geometry}
\usepackage{pgfplots}
\pgfplotsset{compat=newest}
\crefname{equation}{\unskip}{\unskip}

\newcommand{\w}{\boldsymbol{W}}
\newcommand{\x}{\boldsymbol{x}}
\newcommand{\s}{\boldsymbol{s}}
\newcommand{\set}[1]{\mathcal{#1}}

\newcommand*{\Scale}[2][4]{\scalebox{#1}{\ensuremath{#2}}}%

\newtheorem{theorem}{Theorem}

\newtheorem{corollary}{Corollary}
\newtheorem{example}{Example}

\author{\IEEEauthorblockN{Reent Schlegel\IEEEauthorrefmark{1}, Siddhartha Kumar\IEEEauthorrefmark{1}, Eirik Rosnes\IEEEauthorrefmark{1}, and Alexandre Graell i Amat\IEEEauthorrefmark{2}\IEEEauthorrefmark{1}}
\IEEEauthorblockA{\IEEEauthorrefmark{1}Simula UiB, Bergen, Norway
}\IEEEauthorblockA{\IEEEauthorrefmark{2}Department of Electrical Engineering, Chalmers University of Technology, Gothenburg, Sweden
}}
\title{Private Edge Computing for Linear Inference \\Based on Secret Sharing}

\begin{document}

	\maketitle

	\begin{abstract}
		We consider an edge computing scenario where users want to perform a linear computation on local, private data and a network-wide, public matrix. Users offload computations to edge servers located at the edge of the network, but do not want the servers, or any other party with access to the wireless links, to gain any information about their data. We provide a scheme that guarantees information-theoretic user data privacy against an eavesdropper with access to a number of edge servers or their corresponding communication links. The novelty of the proposed scheme lies in the utilization of secret sharing and partial replication to provide privacy, mitigate the effect of straggling servers, and to allow for joint beamforming opportunities in the download phase,  to minimize the overall latency, consisting of upload, computation, and download latencies.
		
	\end{abstract}
	\section{Introduction}
	Edge computing has established itself as a pillar of the 5G mobile network \cite{ETSI} to guarantee very low-latency and high-bandwidth computing services. The key idea is to move the computation power from the cloud closer to where data is generated, by pooling the
    available resources at the network edge.

	Processing data in a distributed
	fashion over a number of edge servers poses significant challenges. In particular, edge servers may fail, be inaccessible, or straggle. The straggler problem has recently been  addressed in the context of distributed computing in data centers (over the cloud), where coding has been shown to be a powerful tool to reduce the computational latency due to straggling servers \cite{Li,Lee,Albin1,Albin2}. The idea is to generate redundant computations by means of an erasure correcting code such that the partial computations of a subset of the servers suffice to complete the whole computation, thus providing resiliency to straggling (and failing) servers. The same concept can be applied in edge computing. In this scenario, besides the computational latency due to straggling servers, the communication latency of uploading and downloading data to the 
	servers is of utmost importance, due to severe bandwidth limitations.

	To reduce the communication latency, in \cite{Tao,TaoStudy} subtasks were replicated across edge servers to enable cooperation opportunities to send results back to the users  via joint  beamforming.
	More recently, \cite{Osvaldo,Kuikui} combined both  straggler coding using a maximum distance separable (MDS) code and joint beamforming  to reduce the overall latency. Another important challenge when processing data over heterogeneous, untrusted edge servers is guaranteeing the privacy of the user data. Recently, this problem has been addressed in the context of distributed computing in data centers in the presence of straggling servers \cite{SalimStair,SalimRateless}. These works use secret sharing ideas to provide both privacy and robustness against stragglers. 
	
	In this paper, we propose a privacy-preserving edge computing scheme that exploits straggler coding and partial replications across servers  to reduce latency.
	To the best of our knowledge, this problem has not been considered before in the literature. In particular, we consider a similar scenario to the one in \cite{Osvaldo} where multiple users wish to perform a linear inference on some local data given a network-wide,  public matrix. Practical examples where such a scenario arises include recommender systems via collaborative filtering.

	For this scenario, we present a scheme that guarantees information-the\-or\-etic user data privacy against an eavesdropper with access to a number of edge servers or their corresponding communication links. The proposed scheme utilizes secret sharing to provide both  privacy and mitigate the effect of straggling servers. Furthermore, by replicating computations across different servers the scheme allows for joint beamforming opportunities. The proposed scheme entails an inherent tradeoff between computational latency due to stragglers, communication latency, and user data privacy. For a given privacy level, we optimize the parameters of the scheme in order to  minimize the overall latency incurred by the upload and  download of data as well as the computation. For the lowest privacy level, i.e., privacy against a single untrusted server, the proposed scheme yields an increase in  latency in the worst case by a moderate factor of about $2.4$ compared to the nonprivate scheme in \cite{Osvaldo} for the selected system parameters.

	\textit{Notation:} Vectors and matrices are written in lowercase and uppercase bold letters, respectively, e.g.,  $\boldsymbol{a}$ and $\boldsymbol{A}$. The transpose of vectors and matrices is denoted by $(\cdot)^\top$. $\text{GF}(q)$ denotes the finite field of order $q$ and $\mathbb{N}$ denotes the positive integers. We use the notation $[a]$ to represent  the set of integers $\{0,1,\ldots,a-1\}$. Furthermore, $\left\lceil a/b\right\rceil$ is the smallest integer larger than or equal to $a/b$, $\left\lfloor a/b\right\rfloor$ is the largest integer smaller than or equal to $a/b$,  and $(a)_b$ is the integer $a$ modulo $b$. We represent permutations in cycle notation, e.g., the permutation $\pi = (0\;2\;1\;3)$ maps $0\mapsto2$, $2\mapsto1$, $1\mapsto3$, and $3\mapsto0$. In addition, $\pi(i)$ is the image of $i$ under $\pi$, e.g., $\pi(0) = 2$. The expected value of a random variable $X$ is denoted by $\mathbb{E}[X]$.

	\section{System Model}\label{Sec: SystemModel}
\begin{figure}
		\centering
		\includegraphics[keepaspectratio, width = 0.7\columnwidth]{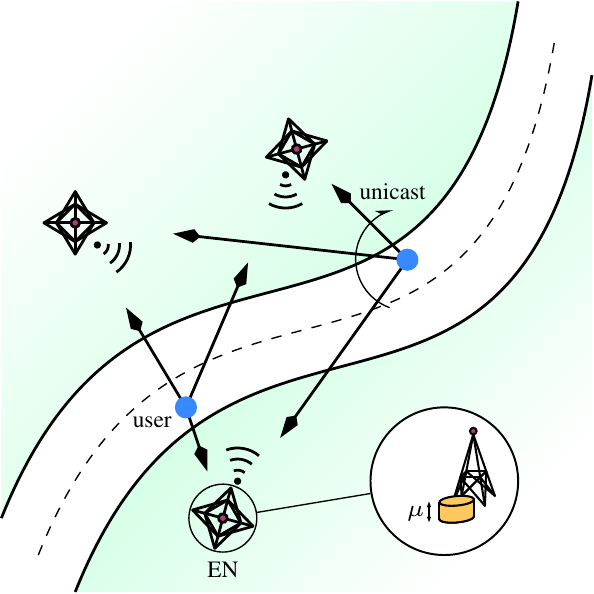}
		\caption{An edge computing network with two users and three ENs.\label{fig:system_model}}
		\vspace{-3ex}
	\end{figure}

	We consider the system in \cref{fig:system_model} with \(u\) users $\mathsf{u}_0, \mathsf{u}_1,\ldots ,\mathsf{u}_{u-1}$, where the data of user $\mathsf{u}_i$ is represented by the vector \(\x_i = (x_{i,0},x_{i,1},\ldots,x_{i,r-1})^\top\in\mathrm{GF}{(q)}^r\). Each user $\mathsf{u}_i$ wants to perform a computation-intensive linear inference \(\w\x_i\), where \(\w\in\textrm{GF}{(q)}^{m\times r}\), in a distributed fashion over \(e\) edge nodes (ENs) $\mathsf{e}_0, \mathsf{e}_1,\ldots,\mathsf{e}_{e-1}$ located at the edge of the network. For ease of notation we will refer to the set $\{\w\x_i\mid i\in [u]\}$ as $\{\w\x_i\}$ and to $\{\x_i\mid i\in [u]\}$ as $\{\x_i\}$. The matrix $\w$ stays constant for a sufficiently long period of time, and each EN has a storage capacity corresponding to a fraction \(\mu\), \(0<\mu \leq 1\), of the matrix \(\w\), which is assumed to be public. Moreover, we assume that each user is connected by \(e\) unicast wireless links to the \(e\) ENs.
	\subsection{Computation Runtime Model}
	The ENs may straggle, which is represented by a random setup time \(\lambda_j\) for each EN \(\mathsf{e}_j\). The setup time is the time it takes an EN to start computing after it has received the necessary data. As in \cite{Osvaldo,Dean,Mallick}, we assume that the setup times are independent and identically distributed (i.i.d.) according to an exponential distribution with parameter \(\eta\), such that $\mathbb{E}[\lambda_j] = 1/\eta$. The time it takes an EN to compute one inner product in \(\text{GF}(q)^r\) for each of the users is deterministic and denoted by \(\tau\). Thereby, $\tau$ captures both the effect of the total number of users $u$ and the computational capabilities of the ENs. Thus, the latency incurred by EN $\mathsf{e}_j$ to compute $d$ inner products for each user ($u\cdot d$ inner products in total) is
	\begin{equation*}
		\mathsf{L}^\mathsf{comp}_{j} = \lambda_{j} + d\tau.
	\end{equation*}
	We define the \emph{normalized computation} latency of EN $\mathsf{e}_j$ as
	\begin{equation*}
		\mathsf{\tilde{L}}^\mathsf{comp}_{j} = \frac{\mathsf{L}^\mathsf{comp}_{j}}{\tau} = \frac{\lambda_{j}}{\tau} + d.
	\end{equation*}
	\subsection{Communication}
	Both the upload of data from the users to the ENs and the download of the results of the computations from the ENs to the users is considered. We denote by $\gamma$ the normalized communication latency of unicasting $u$ symbols from $\text{GF}(q)$ in the upload or download. In the uplink, each user unicasts its data vector for computation to the ENs. In the downlink, ENs having access to the same symbol can collaboratively transmit to multiple users at the same time and thereby reduce the communication latency by exploiting joint beamforming opportunities \cite{Tao,Kuikui,Osvaldo,TaoStudy,Zhang,Naderializadeh}. In particular, a symbol available at $\rho$ ENs can be transmitted simultaneously to $\min\{\rho,u\}$ users with a normalized communication latency of \(\gamma/\min\{\rho,u\}\) in the high signal-to-noise (SNR) region. The normalized communication latency,  in the high SNR region, of transmitting $v$ symbols, where symbol $\alpha_i$, $i\in[v]$, is available at $\rho_i$ ENs, is
	\begin{equation*}
		\mathsf{\tilde{L}}^\mathsf{comm,down} = \gamma \sum_{i=0}^{v-1} \frac{1}{\min\{\rho_i,u\}}.
	\end{equation*}

	\subsection{Privacy and Problem Formulation}

	We consider a scenario where some of the ENs or their corresponding communication links are compromised. In particular, we assume the presence of an eavesdropper with access to any $z$ ENs or their corresponding communication links.
	
	The goal is to offload computations to the honest but curious ENs in such a way that they do not gain any information in an information-theoretic sense (zero mutual information) about neither the user data $\{\x_i\}$ nor the results of the computations $\{\w\x_i\}$, while minimizing the overall normalized latency, consisting of upload, computation, and download latencies.
		
	\section{Private Distributed Linear Inference}
	In this section, we present a distributed linear inference computation scheme that provides user data privacy against an eavesdropper with access to any \(z\) ENs or their corresponding communication links. At the heart of the proposed scheme lies Shamir's secret sharing scheme (SSS) \cite{Shamir}. An SSS with parameters $(n,k)$, $n\geq k$, ensures that some private data can be shared with $n$ parties in such a way that any $k-1$ colluding parties do not learn anything about the data. On the other hand, any set of $k$ or more parties can recover the data.
	
	For each user $\mathsf{u}_i$, Shamir's \((n, k)\) SSS is used to compute \(n\) shares of its private data  \(\x_i = (x_{i,0}, x_{i,1},  \ldots,x_{i,r-1})^\top\). In particular, for user $\mathsf{u}_i$ we encode each data entry \(x_{i,l}\) along with $k-1$ i.i.d.\ uniform random symbols $r_{i,l}^{(1)},r_{i,l}^{(2)},\ldots,r_{i,l}^{(k-1)}$ from $\text{GF}(q)$, where $q > n$, using an \((n,k)\) Reed-Solomon (RS) code  to obtain \(n\) coded symbols  $s_{i,l}^{(0)},s_{i,l}^{(1)},\ldots ,s_{i,l}^{(n-1)}$. For each $h\in[n]$, the $(h+1)$-th share of user $\mathsf{u}_i$ is 
	\begin{align*}
		\s^{(h)}_i=\left(\begin{matrix}
			s^{(h)}_{i,0}, s^{(h)}_{i,1}, \ldots, s^{(h)}_{i,r-1}\\
		\end{matrix}\right)^\top.
	\end{align*}
	Finally, define the matrix of shares
	\begin{align}
	\label{eq:matrix_of_shares}
		\bm S^{(h)}=\left(\begin{matrix}
			\s^{(h)}_0, \s^{(h)}_1, \ldots, \s^{(h)}_{u-1}
		\end{matrix}\right)\in\text{GF}{(q)}^{r\times u}
	\end{align}
	as the matrix collecting the $(h+1)$-th share of all users.

	The following theorem proves that the original computations \(\{\w\x_i\}\) of all users can be recovered from a given  set of computations based on the matrices of shares $\boldsymbol{S}^{(0)},\boldsymbol{S}^{(1)},\ldots, \boldsymbol{S}^{(n-1)}$,
	while  providing  privacy against an eavesdropper with access to at most \(k-1\) distinct matrices of shares.

	\begin{theorem}\label{Th: RecoverSecretComputation}
		Consider \(u\) users with their respective private data $\x_i\in\text{GF}(q)^r$, $i\in[u]$. Use Shamir's \((n,k)\) SSS on each \(\x_i\) to obtain the matrices of shares \(\boldsymbol{S}^{(0)},\boldsymbol{S}^{(1)},\ldots, \boldsymbol{S}^{(n-1)}\) in \eqref{eq:matrix_of_shares}. 
		Let \(\w\in\text{GF}{(q)}^{m\times r}\) be a public matrix and \(\mathcal I\subseteq[n]\) a set of indices with cardinality \(|\mathcal I|=k\). Then, the set of computations \(\{\w\bm S^{(h)}\mid h\in\mathcal I\}\) allows to recover the computations \(\{\w\x_i\}\) of all users. Moreover, for any set $\mathcal{J}\subseteq[n]$ with $|\mathcal{J}| < k$, \(\{\w\bm S^{(h)}\mid h\in\mathcal J\}\) reveals no information about  \(\{\w\x_i\}\).
	\end{theorem}
	\begin{IEEEproof}
		Let \(\set{C}\) be the $(n,k)$ RS code used in the SSS. For each \(h\in[n]\), the entries of the rows of \(\bm S^{(h)}\) are code symbols in position \(h\) of codewords from \(\set{C}\) pertaining to different users. More precisely, for each user $\mathsf{u}_i$, each row of the matrix \(\bigl(\s^{(0)}_i, \s^{(1)}_i, \ldots,\s^{(n-1)}_{i}\bigr)\) of all $n$ shares of $\mathsf{u}_i$ is a codeword from \(\set{C}\). Since \(\set{C}\) is a linear code, each of the $m$ rows of the matrix 
		\begin{align*}
			\w\left(\begin{matrix}
				\s^{(0)}_i, \s^{(1)}_i,\ldots, \s^{(n-1)}_{i}
			\end{matrix}\right)
		\end{align*}
		is a codeword of \(\set{C}\). Furthermore, the messages obtained by decoding these codewords are the rows of
		\begin{align*}
			\left(\w\x_i,\w\bm r_i^{(1)},\ldots,\w\bm r_i^{(k-1)}\right),
		\end{align*}
		where $\{ \bm r_i^{(\kappa)}=(r^{(\kappa)}_{i,0},r^{(\kappa)}_{i,1},\ldots,r^{(\kappa)}_{i,r-1})^\top \mid\kappa \in [k]\backslash\{0\}\}$ is the set of  vectors of uniform random symbols used by user $\mathsf{u}_i$ in the computation of the shares $\bm s_i^{(h)}$, $h \in [n]$. Then, decoding the vectors in the set
		\(\{\w\s^{(h)}_i\mid h\in\set{I}\}\)
		gives \(\w\x_i\), and it follows that \(\{\w\bm S^{(h)}\mid h\in\mathcal I\}\) gives  \(\{\w\x_i\}\).

		From the properties of Shamir's SSS it follows  that the mutual information between \(\{\bm S^{(h)}\mid h\in\mathcal J\}\) and $\{\x_i\}$ is zero. Subsequently, from the data processing inequality it follows  that \(\{\w\bm S^{(h)}\mid h\in\mathcal J\}\) reveals no information about $\{\x_i\}$.
	\end{IEEEproof}
	The following corollary gives a sufficient condition to recover the private computations \(\{\w\x_i\}\).
	\begin{corollary}[Sufficient recovery condition]\label{Cor: Recovery}
		Consider an edge computing scenario, where the public matrix \(\w\) is partitioned into \(b\) disjoint submatrices \(\w_l\in\text{GF}{(q)}^{\frac{m}{b}\times r}\), $l\in[b]$, and the private data is \(\{\x_i \}\). Then, the private computations \(\{\w\x_i \}\) can be recovered from the computations in the sets 
		\begin{align} 
		\label{Eq: RecoveryCondition}	
			\set{S}_l\triangleq\{\w_l\bm S^{(h)}\mid h\in\mathcal I\},\; l\in[b],
		\end{align} 
		for any fixed set $\mathcal{I}\subseteq[n]$ with cardinality $|\mathcal{I}| = k$.  
	\end{corollary}
	\begin{IEEEproof}
		From \cref{Th: RecoverSecretComputation}, for a given \(l\in[b]\),  the computations in the set $\{\w_l\x_i\}$ can be recovered from the computations in the set $\set{S}_l$. Then, we obtain 
		\[\w\x_i=\left(\begin{matrix}
			(\w_0\x_i)^\top, (\w_1\x_i)^\top, \ldots, (\w_{b-1}\x_i)^\top
		\end{matrix}\right)^\top, \forall\, i\in[u].\]
	\end{IEEEproof}
	
	In the following, we present a scheme that fulfills the sufficient recovery condition in \cref{Cor: Recovery}. Note that it may be beneficial to repeat shares over several ENs in order to exploit broadcasting opportunities during the download phase. This presents difficulties in the design of a private scheme, because repeating shares at different nodes results in a privacy level $z$ lower than that of the SSS ($k$). For example, if all ENs have access to two matrices of shares, the scheme only provides privacy against any $z=\lfloor (k-1)/2 \rfloor$ colluding ENs.

	Given the underlying SSS, the proposed scheme can be broken down into two combinatorial problems. The first corresponds to the assignment of the submatrices \(\{\w_l \mid l \in [b]\}\) to the \(e\) ENs such that no EN stores more than a fraction $\mu$ of $\w$. The second corresponds to the assignment of the $n$ matrices of shares $\{\bm S^{(h)} \mid h \in [n]\}$ to the ENs such  that the users are guaranteed to obtain the computations in \cref{Eq: RecoveryCondition}.
	  
	\subsection{Assignment of \(\w\) to the Edge Nodes}\label{Sec: Wassignment}
		We start by explaining the assignment of the submatrices of $\w$ to the ENs such that no EN stores more than a fraction $\mu$ of $\w$, while the users are guaranteed to recover their computations $\{\w \x_i\}$. Additionally, we would like to allow for replications across different ENs to allow for joint beamforming in the download phase.

		In order to satisfy the storage requirement, we select \(p\in\mathbb{N}\) such that \(p/e \leq \mu\) and partition \(\w\) into \(b=e\) submatrices as
		\begin{align*}
			\w=\left(\begin{matrix}
				\w_0^\top, \w_1^\top, \ldots, \w_{e-1}^\top
			\end{matrix}\right)^\top.	
			\end{align*}
		We then assign \(p\) submatrices to each of the \(e\) ENs. The assignment has the following combinatorial structure. Consider a cyclic permutation group of order \(e\) with generator \(\pi\). We construct an index matrix 
		\begin{align}
			\label{Eq: Iw}
			\bm{I}_{\mathsf w} &\triangleq \left(\begin{matrix}
			\pi^0(0) & \pi^0(1) & \cdots & \pi^0(e-1)\\
			\pi^1(0) & \pi^1(1) & \cdots & \pi^1(e-1)\\
			\vdots & \vdots & \ddots & \vdots\\
			\pi^{p-1}(0) & \pi^{p-1}(1) & \cdots & \pi^{p-1}(e-1)\\
			\end{matrix}\right)
		\end{align}
		and define the set of indices
		\begin{equation}
			\label{Eq:setIw}
			\set{I}_j^{\mathsf w}=\{\pi^{0}(j),\pi^{1}(j),\ldots,\pi^{p-1}(j)\}
		\end{equation}
		for $j\in[e]$ as the set containing the elements in column \(j\) of \(\bm I_{\mathsf w}\). Then, we assign the submatrices \(\{\w_l\mid l\in\set{I}_j^{\mathsf w}\}\) to EN $\mathsf{e}_j$. For example, if \(\pi=(0\;e-1\; e-2\; \cdots\; 1)\), we have 
		\begin{align*}
			\bm{I}_{\mathsf w}=\left(\begin{matrix}
			0 & 1 & \cdots & e-1\\
			e-1 & 0 & \cdots & e-2\\
			\vdots & \vdots & \ddots & \vdots\\
			e-p+1 & e-p+2 & \cdots & e-p
			\end{matrix}\right),
		\end{align*}
		and EN $\mathsf{e}_1$ stores $\w_1, \w_{0}, \w_{e-1},\ldots,\w_{e-p+2}$. 
		
		The ENs process the assigned submatrices of $\w$ in the same order as their indices appear in the rows of $\bm{I}_{\mathsf{w}}$, and we define $\phi_j^\mathsf{w}(l)$ for $l\in [p]$ to be the map to the index of the $(l+1)$-th assigned submatrix of EN $\mathsf{e}_j$.

	\subsection{Assignment of Shares to the Edge Nodes}

	Given the assignment of the submatrices of \(\w\), we now have to assign the shares in such a way that we can guarantee that the users obtain the computations in \cref{Eq: RecoveryCondition}. The users upload their shares to the \(e\) ENs according to the following assignment. Given the generator \(\pi\) used to assign the submatrices of \(\w\) to the ENs, we construct a \((\beta+1)\times e\) index matrix
	\begin{align}
		\label{Eq: Is}
		\bm I_{\mathsf s}=\Scale[0.972]{\left(\begin{matrix}
			\pi^{0}(0) & \pi^{0}(1) & \cdots & \pi^{0}(e-1)\\
			\pi^{e-p}(0) & \pi^{e-p}(1) & \cdots & \pi^{e-p}(e-1)\\
			\vdots & \vdots & \ddots & \vdots\\
			\pi^{\beta(e-p)}(0) & \pi^{\beta(e-p)}(1) & \cdots & \pi^{\beta(e-p)}(e-1)
		\end{matrix}\right)},
	\end{align}
	where \(\beta=\left\lceil e/p\right\rceil-1\). Define the set of indices
	\begin{equation}
		\label{Eq:setIs}
		\mathcal{I}_j^{\mathsf{s}}=\{\pi^{0}(j),\pi^{1}(j),\ldots,\pi^{\beta(e-p)}(j)\}\backslash\{n, n+1,\ldots,e-1\}
	\end{equation}
	as the subset of elements in column \(j\) of \(\bm I_{\mathsf s}\) that are in $[n]$.  User $\mathsf{u}_i$ transmits the shares \(\{\bm s_i^{(h)}\mid h\in\mathcal{I}_j^{\mathsf{s}}\}\) to EN \(\mathsf{e}_j\). In case $|\mathcal{I}_j^{\mathsf{s}}| <  a$, where $a=\left\lceil \lceil e/p\rceil \cdot n/e\right\rceil$, for some $j$, additional rows are added to $\bm I_{\mathsf s}$ in order to fill up these  sets such that $|\mathcal{I}_j^{\mathsf{s}}| = a$, $\forall\,  j\in[e]$.	These combined assignments of submatrices and shares to the ENs allow all users to obtain enough partial computations from the \(e\) ENs to retrieve their desired computations, as will be shown in \cref{Th: recover}.
	As for the submatrices of $\w$, the shares are processed in the same order as their indices appear in the rows of $\bm{I}_{\mathsf{s}}$, and we define $\phi_j^\mathsf{s}(h)$ for $h\in [a]$ to be the map to the index of the $(h+1)$-th assigned matrix of shares of EN $\mathsf{e}_j$. For a given matrix of shares assigned to an EN, all assigned submatrices of $\w$ are processed before moving on to the next matrix of shares.
	
	\begin{theorem}\label{Th: recover}
	Consider an edge computing network consisting of \(u\) users and \(e\) ENs, each with a storage capacity corresponding to a fraction  \(\mu\), $0 < \mu \leq 1$, of $\w$, and Shamir's  SSS with \(n\leq e\) shares. For \(j\in[e]\), EN \(\mathsf{e}_j\) stores the submatrices of \(\w\) from the set \(\{\w_l\mid l\in\set{I}_j^{\mathsf w}\}\) with $\set{I}_j^{\mathsf w}$ defined in \cref{Eq:setIw}. Furthermore, it receives the matrices of  shares from the set  \(\{\bm S^{(h)}\mid h\in\mathcal{I}_j^{\mathsf{s}}\}\) with $\mathcal{I}_j^{\mathsf{s}}$ defined in \cref{Eq:setIs}, and computes and returns the set \(\{\w_l\bm S^{(h)} \mid l\in\set{I}_j^{\mathsf w},  h\in\mathcal{I}_j^{\mathsf{s}} \}\) to the users. Then, all users can recover their desired computations $\{\w \bm x_i \}$.
	\end{theorem}
	Due to lack of space, we omit the proof of \cref{Th: recover}. We motivate the theorem, however, with the following example.
	\begin{example}
		Consider \(e=n=5\), \(p=3\), and \(\pi=(0\;3\;1\;4\;2)\), the generator of a cyclic permutation group of order \(5\). From \cref{Eq: Iw,Eq: Is}, we have
		\begin{align*}
			\bm I_{\mathsf w}&=\left(\begin{matrix}
				0 & 1 & 2 & 3 & 4\\
				3 & 4 & 0 & 1 & 2\\
				1 & 2 & 3 & 4 & 0
			\end{matrix}\right) \text{ and }
			\bm I_{\mathsf s}&=\left(\begin{matrix}
				0 & 1 & 2 & 3 & 4\\
				1 & 2 & 3 & 4 & 0
			\end{matrix}\right).
		\end{align*}
		We focus on the matrix  of shares $\bm S^{(0)}$. It is assigned to EN $\mathsf{e}_0$ and gets multiplied with the submatrices of $\bm W$ indexed by the elements of the set
		\begin{align*}
		   \set{I}_0^{\mathsf w} = \{\pi^0(0), \pi(0), \pi^2(0)\}=\{0,3,1\}.
		\end{align*}
		Note that the set $\set{I}_0^{\mathsf w}$ contains three recursively $\pi$-permuted integers of $0$  ($\pi^0(0)$, $\pi^1(0)$, and $\pi^2(0)$). Now, consider EN $\mathsf{e}_4$, which is also assigned the matrix  of shares $\bm S^{(0)}$. We have
		\begin{align*}
			\set{I}_4^{\mathsf w} = \{\pi^0(4), \pi(4), \pi^2(4)\}=\{4,2,0\}.
		\end{align*}
		Notice that \(\pi^0(4) = \pi^3(0)=4\) is the fourth recursively $\pi$-permuted integer of $0$. 
		Hence, the set \(\set{I}_0^{\mathsf w}\cup\set{I}_4^{\mathsf w}\) contains in total six recursively $\pi$-permuted integers of $0$, which is sufficient to give the set \([5]\), since the group generated by $\pi$ is transitive. In a similar way, it can be shown that the same property holds for all other matrices of shares. Therefore, each matrix of shares is multiplied with all submatrices of $\bm W$, and the sets in \cref{Eq: RecoveryCondition} are obtained.
	\end{example}

	\section{Communication and Computation Scheduling}
	In this section, we describe the scheduling of uploading the assigned shares to the ENs, performing the computations, and downloading a subset of $\{\w_l\boldsymbol{S}^{(h)} \mid l \in \mathcal{I}_j^{\mathsf{w}}, h\in\mathcal{I}_j^{\mathsf{s}},j \in [e]\}$. In the following, we refer to a single $\w_l\boldsymbol{S}^{(h)}$ as an intermediate result (IR).
	\subsection{Upload and Computation}
	As $\w$ stays constant for a long time, the assignment of the submatrices $\{\w_l \mid l \in [e] \}$ can be done offline and does not affect the overall latency. The online phase starts with the upload of the shares. In contrast to the nonprivate scheme in \cite{Osvaldo}, a user $\mathsf{u}_i$ can not broadcast one vector to all ENs. Instead, the user has to unicast a number of shares to each EN to assure that any $z$ ENs do not obtain any information about $\x_i$. In general, broadcasting a message to $e$ receivers is more expensive than transmitting a single unicast message to one receiver. As in \cite{Lee}, we assume that broadcasting to $e$ receivers is a factor $\log(e)$ more expensive in terms of latency than a single unicast. Recall that the cost (or normalized latency) of unicasting $u$ symbols from $\text{GF}(q)$ is $\gamma$. Hence, in the nonprivate scheme the normalized latency of every user broadcasting one vector from $\text{GF}(q)^r$ to all $e$ ENs is $\mathsf{\tilde{L}}^\mathsf{up}_\mathsf{NP} = \gamma \cdot r\cdot \log(e)$.

	\begin{figure}
		\includegraphics[width=\columnwidth]{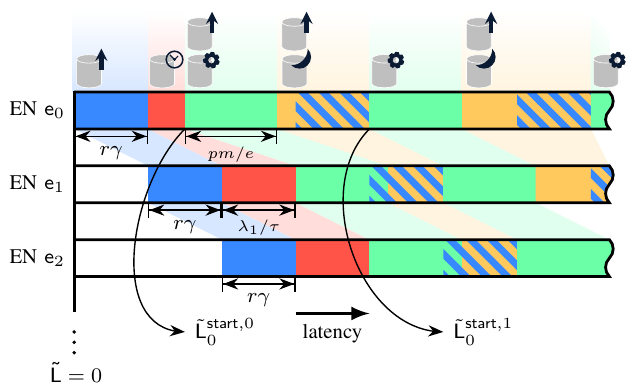}
		\vspace{-3ex}
		\caption{Upload and computing schedule. For each EN $\mathsf{e}_j$, the upload times $r\gamma$ are shown in blue, the random setup times in red, the times $pm/e$ to compute $p$ IRs in green, and  possible idle times in yellow. All times are normalized.}
		\label{fig:schedule}
		\vspace{-2ex}
	\end{figure}

	In contrast, the normalized latency of unicasting $u$ shares, one from each user, which are elements in $\text{GF}(q)^r$, to one EN is $\gamma r$. Recall that each EN receives $a$ matrices of shares. We assume that each user can upload only one share to one EN at a time. The upload is illustrated in \cref{fig:schedule}, in which the blue segments correspond to the upload phase. We start by uploading the first matrix of shares to EN $\mathsf{e}_0$, continue with EN $\mathsf{e}_1$, and proceed until all ENs have received their first matrix of shares. This process is repeated with the remaining matrices of shares until EN $\mathsf{e}_j$ has received the $a$ matrices of shares $\{\boldsymbol{S}^{(h)} \mid h\in\mathcal{I}_j^{\mathsf{s}}  \}$,  $j \in [e]$. EN $\mathsf{e}_j$ receives its $(h+1)$-th matrix of shares $\bm S^{(\phi^{\mathsf{s}}_j(h))}$ at normalized time
	\begin{equation*}
		\mathsf{\tilde{L}}^{\mathsf{up},h}_j =  \gamma r(eh + j+1),
	\end{equation*}
	and the total normalized upload latency of the private scheme becomes $\mathsf{\tilde{L}}^\mathsf{up}_\mathsf{P} = \gamma \cdot r\cdot e\cdot a$.

	After an EN has received its first matrix of shares, it enters the computation phase. As mentioned earlier, the ENs experience a random setup time before they can start their computations. This is illustrated by the red segments in \cref{fig:schedule}. For EN $\mathsf{e}_j$ this phase incurs a normalized latency of $\lambda_j/\tau$. Once set up, the ENs start their computations on the first assigned matrix of shares. In total, $p$ IRs of the form $\w_l\bm S^{(h)}$ have to be computed for each assigned matrix of shares $\bm S^{(h)}$ by EN $\mathsf{e}_j$, where  $l \in  \set{I}_j^{\mathsf w}$ and $h \in  \set{I}_j^{\mathsf s}$. This incurs a normalized latency of $p\cdot m/e$, because each $\w_l$ has $m/e$ rows and hence, the ENs compute $u\cdot m/e$ inner products for each of the $p$ IRs.

	In the case an EN has not received another matrix of shares before finishing the currently assigned computations, it remains idle until it receives another matrix of shares to compute on. This can be seen in yellow in \cref{fig:schedule}. 
	For $h \in [a]$, the normalized time at which EN $\mathsf{e}_j$ starts to compute on the $(h+1)$-th assigned matrix of shares, i.e., on $\bm S^{(\phi^{\mathsf{s}}_j(h))}$, is 
	\begin{equation*}
		\mathsf{\tilde{L}}^{\mathsf{start},h}_j = \max\left \{\mathsf{\tilde{L}}^{\mathsf{start},h-1}_j + p\frac{m}{e}~,~\mathsf{\tilde{L}}^{\mathsf{up},h}_j\right\},\; \text{for $h > 0$},
	\end{equation*}
	with
	\begin{equation*}
		\mathsf{\tilde{L}}^{\mathsf{start},0}_j = \frac{\lambda_j}{\tau} + \mathsf{\tilde{L}}^{\mathsf{up},0}_j.
	\end{equation*}
	The computational phase continues at least until the computations in \cref{Eq: RecoveryCondition} are obtained, i.e.,  until there are at least $k$ distinct IRs of the form $\w_l \bm S^{(h)}$, $h \in [n]$, for each $l \in [e]$. This ensures that a given user $\mathsf{u}_i$ can recover  \(\w\x_i\). It can be beneficial to continue computing products to reduce the communication latency in the download phase, as we discuss next.
	
	\subsection{Download}
	In the download phase we can make use of joint beamforming opportunities to reduce the latency by serving multiple users at the same time. An IR $\w_l\boldsymbol{S}^{(h)}$ that is computed at $\rho_{l,h}$ ENs incurs a normalized communication latency of $\gamma /\min\{\rho_{l,h},u\}$. Hence, a higher multiplicity of computed IRs across different ENs will reduce the communication latency in the download phase. At the same time, the repeated IRs have to be computed first, thereby increasing the computational latency. This tradeoff can be optimized to reduce the overall latency. Assume the optimum is reached after EN $\mathsf{e}_{j^*}$ has computed the IR $\w_{\phi^\mathsf{w}_j(l^*)}\boldsymbol{S}^{(\phi^\mathsf{s}_j(h^*))}$. This gives a normalized computation latency of 
	\begin{equation*}
		\mathsf{\tilde{L}}^\mathsf{comp} = \mathsf{\tilde{L}}^{\mathsf{start},h^*}_{j^*} + (l^*+1) \frac{m}{e}.
	\end{equation*}

	After the computation phase has finished, the ENs cooperatively send the computed IRs $\w_l\boldsymbol{S}^{(h)}$ simultaneously to multiple users in descending order of their multiplicity $\rho_{l,h}$ until the computations in \cref{Eq: RecoveryCondition} are available to the users. More precisely, for each $\w_l$ the ENs send the $k$ IRs with the highest multiplicities to the users. Then, a given user $\mathsf{u}_i$ can decode the SSS to obtain the desired computation $\w\x_i$. For a fixed $l$, let $\mathcal{H}_l^\mathsf{max} = \arg\max_{\mathcal{A}\subseteq[n],|\mathcal{A}| = k} \sum_{h\in\mathcal{A}} \rho_{l,h}$ be the set of indices $h$ of the $k$ largest $\rho_{l,h}$. This results in a normalized communication latency   of
	\begin{equation*}
		\mathsf{\tilde{L}}^\mathsf{comm} = \gamma \sum_{l = 0}^{e-1}\sum_{h\in \mathcal{H}_l^\mathsf{max}} \frac{1}{\min\{\rho_{l,h},u\}},
	\end{equation*}
	and the overall normalized latency becomes
	\begin{equation}
	\label{eq:overall_latency}
		\mathsf{\tilde{L}} = \mathsf{\tilde{L}}^{\mathsf{start},h^*}_{j^*} + (l^*+1) \frac{m}{e} + \gamma \sum_{l = 0}^{e-1}\sum_{h\in \mathcal{H}_l^\mathsf{max}} \frac{1}{\min\{\rho_{l,h},u\}}.
	\end{equation}

	\section{Optimization and Numerical Results}
	We start by explaining how to choose the parameters of the proposed scheme so that the overall normalized latency $\mathsf{\tilde{L}}$ in \eqref{eq:overall_latency}, consisting of upload, computation, and download latencies, is minimized for a given privacy level $z$. To reduce the upload latency, it may be beneficial to contact fewer ENs than the maximum number of ENs available, denoted by \({e}_{\max}\), to which a user can connect. Additionally, storing fewer than \(\mu e\) submatrices of \(\w\) at the ENs can be advantageous, because the ENs will start computations sooner on the later shares. Thus, we can choose \(p \leq \mu e\). 
	
	From the combinatorial designs, it follows that the number of shares $n$ per user can be at most equal to $e$, while the value of the SSS threshold $k$ is constrained by the choices of \(z\), \(e\), \(n\), and \(p\). First, recall that the total number of shares per user assigned to each EN is \(a = \left\lceil \lceil e/p\rceil \cdot n/e\right\rceil\), which means that any $z$ ENs have access to $a\cdot z$ possibly distinct shares of each user. Given that this set of shares must not leak any information about the private data $\{\x_i\}$, we have to pick \(k \geq a z + 1\). According to \cref{Cor: Recovery}, for a given \(\w_l\), waiting for \(k\) distinct products allows to recover the computation $\w_l\x_i$ for each user $\mathsf{u}_i$. Note that there  is no reason to pick \(k\) larger than $a z +1$, since then the users have to wait for more products, leading to reduced straggler mitigation  and increased computational latency. Therefore, we set \(k = a z + 1\). Finally, we need to verify that all  constraints on $n$ are fulfilled, i.e.,  $k\leq n\leq e$ (for the scheme to be feasible), $n\geq k$ (for the SSS to work), and $n\leq e$ (from the combinatorial designs).
 
	We have chosen $\pi = (0\;e-1\;e-2\;\cdots\;1)$ and performed an exhaustive search for the minimum expected overall normalized latency $\mathsf{\tilde{L}}$ given in \eqref{eq:overall_latency} over all valid parameter tuples \((e, n, p)\) for a given privacy level $z$. For each tuple we varied the number of total (not necessarily distinct) IRs to wait for across all ENs for each $\w_l$, in order to minimize the latency. We  generated $10^6$ instances of the random setup times $\{\lambda_j\}$ in the simulation of the scheme in order to obtain an accurate estimate of the expected overall normalized latency.
	
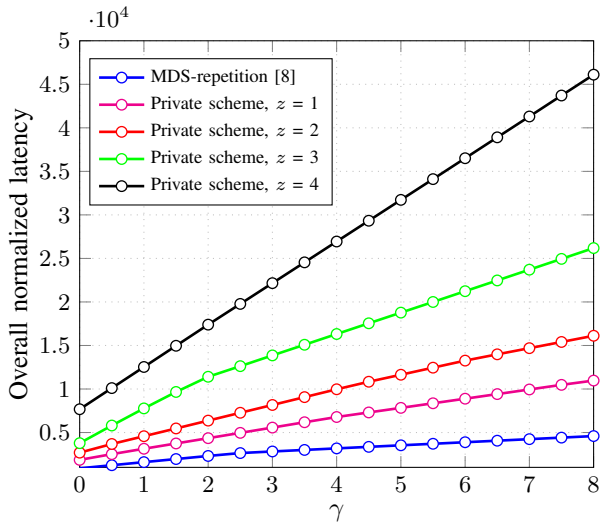
\begin{figure}
		\centering
	  	\begin{tikzpicture}
	  
			\pgfplotsset{every tick label/.append style={font=\small}}
			\begin{axis}[
				width=0.95\columnwidth,
				xmin=0,
				xmax=8, 
				xlabel={\(\gamma\)},
				xlabel style={
						yshift=0.5ex,
						name=label},
				grid style={gray,opacity=0.5,dotted},
				xmajorgrids,
				ymajorgrids,
				ymin=1000, 
				max space between ticks=20pt,
				ymax=50000,
				ylabel={Overall normalized latency},
				ylabel style={
						yshift=-1.0ex,
						name=label},
				axis background/.style={fill=white},
				legend cell align=left,
				legend style={font=\scriptsize, at={(axis cs: 0.15,48000)}, anchor=north west},
				]
				
				\addplot[color=blue,solid,line width=1pt, mark=*, mark options={line width = 0.5pt, fill=white}]table[x=gamma, y=cost]{data/data_MDS_hybrid_9_0.67_0.0005_0.8_600_50.txt};
				\addlegendentry{MDS-repetition \cite{Osvaldo}};

				\addplot[color=magenta,solid,line width=1pt, mark=*, mark options={line width = 0.5pt, fill=white}]table [x=gamma,y=cost]{data/data_private_num_prods_corrected_1000000_1_9_0.6667_0.0005_0.8_600_50.txt};
				\addlegendentry{Private scheme, $z$ = 1};

				\addplot[color=red,solid,line width=1pt, mark=*, mark options={line width = 0.5pt, fill=white}]table [x=gamma,y=cost]{data/data_private_num_prods_corrected_1000000_2_9_0.6667_0.0005_0.8_600_50.txt};
				\addlegendentry{Private scheme, $z$ = 2};

				\addplot[color=green,solid,line width=1pt, mark=*, mark options={line width = 0.5pt, fill=white}]table [x=gamma,y=cost]{data/data_private_num_prods_corrected_1000000_3_9_0.6667_0.0005_0.8_600_50.txt};
				\addlegendentry{Private scheme, $z$ = 3};

				\addplot[color=black,solid,line width=1pt, mark=*, mark options={line width = 0.5pt, fill=white}]table [x=gamma,y=cost]{data/data_private_num_prods_corrected_1000000_4_9_0.6667_0.0005_0.8_600_50.txt};
				\addlegendentry{Private scheme, $z$ = 4};
			\end{axis}
		\end{tikzpicture}
		\vspace{-2ex}
		\caption{Overall normalized latency as a function of $\gamma$ for different privacy levels $z$ of the proposed scheme compared to the nonprivate MDS-repetition scheme in \cite{Osvaldo}. The parameters are \(\mu = 2/3\), \(\tau = 0.0005\), \(\eta = 0.8\), \(e_{\max}=9\), \(m=600\), and \(r=50\).\label{fig:plot_9}}
		\vspace{-3ex}
	\end{figure}

	In \cref{fig:plot_9}, we compare the expected overall normalized latency of the proposed private scheme with the nonprivate MDS-repetition scheme in \cite{Osvaldo}. We plot the overall normalized latency versus $\gamma$ for different privacy levels $z$. For the presented scenario, the users have access to $e_{\max} = 9$ ENs, which can store up to a fraction of $\mu = 2/3$ of the matrix $\w$ with dimensions $m=600$ and $r=50$. The ENs need $\tau = 0.0005$ time units to compute one inner product over $\text{GF}(q)^{50}$ for each of the users, and the straggling parameter is set to $\eta = 0.8$. Providing privacy against a single EN ($z=1$) yields an increase in  latency for $\gamma=8$ by a factor of about $2.4$ compared to the nonprivate MDS-repetition scheme in \cite{Osvaldo}. For $z=2$, the latency increases by a factor of about $3.5$, while it increases to $5.7$ and $10.0$ for $z=3$ and $4$, respectively.

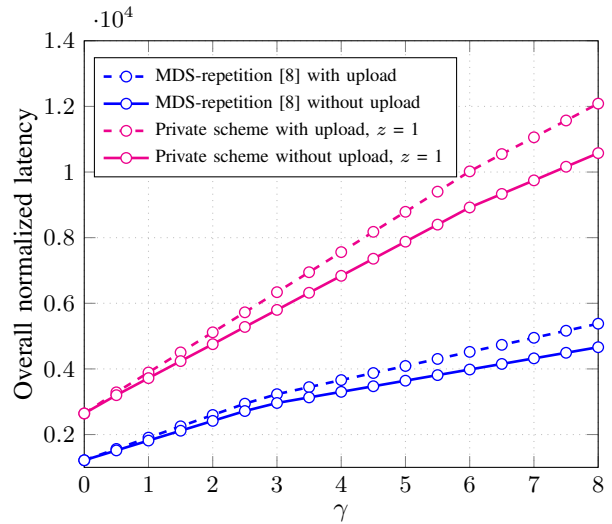
\begin{figure}
		\centering
	  	\begin{tikzpicture}
	  
			\pgfplotsset{every tick label/.append style={font=\small}}
			\begin{axis}[
				width=0.95\columnwidth,
				xmin=0,
				xmax=8, 
				xlabel={\(\gamma\)},
				xlabel style={
						yshift=0.5ex,
						name=label},
				grid style={gray,opacity=0.5,dotted},
				xmajorgrids,
				ymajorgrids,
				ymin=1000, 
				max space between ticks=20pt,
				ymax=14000,
				ylabel={Overall normalized latency},
				ylabel style={
						yshift=-1.0ex,
						name=label},
				axis background/.style={fill=white},
				legend cell align=left,
				legend style={font=\scriptsize, at={(axis cs: 0.15,13500)}, anchor=north west},
				]
				
				\addplot[color=blue,dashed,line width=1pt, mark=*, mark options={line width = 0.5pt, fill=white, solid}]table[x=gamma, y=cost]{data/data_MDS_hybrid_6_0.67_0.0005_0.8_600_50.txt};
				\addlegendentry{MDS-repetition \cite{Osvaldo} with upload};

				\addplot[color=blue,solid,line width=1pt, mark=*, mark options={line width = 0.5pt, fill=white, solid}]table[x=gamma, y=cost]{data/data_MDS_hybrid_6_0.67_0.0005_0.8_600_0.txt};
				\addlegendentry{MDS-repetition \cite{Osvaldo} without upload};

				\addplot[color=magenta,dashed,line width=1pt, mark=*, mark options={line width = 0.5pt, fill=white, solid}]table [x=gamma,y=cost]{data/data_private_num_prods_corrected_1000000_1_6_0.6667_0.0005_0.8_600_50.txt};
				\addlegendentry{Private scheme with upload, $z$ = 1 };

				\addplot[color=magenta,solid,line width=1pt, mark=*, mark options={line width = 0.5pt, fill=white, solid}]table [x=gamma,y=cost]{data/data_private_num_prods_corrected_1000000_1_6_0.6667_0.0005_0.8_600_0.txt};
				\addlegendentry{Private scheme without upload, $z$ = 1};

			\end{axis}
		\end{tikzpicture}%
		\vspace{-2ex}
		\caption{Overall normalized latency as a function of $\gamma$ with and without upload latencies of the proposed scheme compared to the nonprivate MDS-repetition scheme in \cite{Osvaldo}. The parameters are \(\mu = 2/3\), \(\tau = 0.0005\), \(\eta = 0.8\), \(e_{\max}=6\), \(m=600\), and \(r=50\).\label{fig:upload_cost}}
		\vspace{-3ex}
	\end{figure}

	One of the factors that lead to an increased latency is the upload. In the nonprivate scheme, the users can broadcast their data vectors to all ENs simultaneously, whereas in the private scheme, the users have to unicast their shares to the ENs sequentially. In \cref{fig:upload_cost}, we show the impact of the upload on the proposed private scheme and the nonprivate MDS-repetition scheme in \cite{Osvaldo}. At $\gamma = 8$, the upload takes about $13\%$ of the overall latency for both schemes, which yields a latency increase of around $1500$ time units for the private scheme, whereas for the nonprivate scheme it increases by only $700$.
	
\vspace{-1ex}
	\section{Conclusion}
	We presented a privacy-preserving scheme that allows multiple users in an edge computing network to offload computations to edge servers for distributed linear inference, while keeping their data private to a number of edge servers or their corresponding communication links. The proposed scheme uses secret sharing to provide user data privacy and mitigate the effect of straggling servers, and partial repetitions to enable joint beamforming in the download phase  in order to reduce the communication latency. The parameters of the scheme were optimized in order to minimize the overall latency incurred by the upload of data to the servers, the computation, and the transmission of partial computations back to the users.


\begin{thebibliography}{10}
		\providecommand{\url}[1]{#1}
		\csname url@samestyle\endcsname
		\providecommand{\newblock}{\relax}
		\providecommand{\bibinfo}[2]{#2}
		\providecommand{\BIBentrySTDinterwordspacing}{\spaceskip=0pt\relax}
		\providecommand{\BIBentryALTinterwordstretchfactor}{4}
		\providecommand{\BIBentryALTinterwordspacing}{\spaceskip=\fontdimen2\font plus
		\BIBentryALTinterwordstretchfactor\fontdimen3\font minus
		  \fontdimen4\font\relax}
		\providecommand{\BIBforeignlanguage}[2]{{%
		\expandafter\ifx\csname l@#1\endcsname\relax
		\typeout{** WARNING: IEEEtran.bst: No hyphenation pattern has been}%
		\typeout{** loaded for the language `#1'. Using the pattern for}%
		\typeout{** the default language instead.}%
		\else
		\language=\csname l@#1\endcsname
		\fi
		#2}}
		\providecommand{\BIBdecl}{\relax}
		\BIBdecl
		
		\bibitem{ETSI}
		Y.~C. Hu, M.~Patel, D.~Sabella, N.~Sprecher, and V.~Young, ``Mobile edge
		  computing - a key technology towards 5G,'' \emph{ETSI white paper}, no.~11, pp. 1--16, Sep. 2015.

		\bibitem{Lee}
		K.~{Lee}, M.~{Lam}, R.~{Pedarsani}, D.~{Papailiopoulos}, and K.~{Ramchandran},
		  ``Speeding up distributed machine learning using codes,'' \emph{IEEE Trans. Inf. Theory}, vol.~64, no.~3, pp. 1514--1529, Mar. 2018.
		  
		\bibitem{Li}
		S.~{Li}, M.~A. {Maddah-Ali}, and A.~S. {Avestimehr}, ``A unified coding
		  framework for distributed computing with straggling servers,'' in \emph{Proc. IEEE Globecom Workshops (GC Wkshps)}, Washington, DC, Dec. 2016.
		

		
		\bibitem{Albin1}
		A.~{Severinson}, A.~{Graell i Amat}, and E.~{Rosnes}, ``Block-diagonal and LT codes for distributed computing with straggling servers," \emph{IEEE Trans. Commun.}, vol.~67, no.~3, pp. 1739--1753, Mar. 2019.

		\bibitem{Albin2}
		A.~{Severinson}, A.~{Graell i Amat}, E.~{Rosnes}, F.~{Lázaro}, and G.~{Liva}, ``A droplet approach based on Raptor codes for distributed computing with straggling servers," in \emph{Proc. Int. Symp. Turbo Codes 
		Iterative Inf. Processing (ISTC)}, Hong Kong, China, Dec. 2018.

		\bibitem{Tao}
		K.~{Li}, M.~{Tao}, and Z.~{Chen},  ``Exploiting computation replication for mobile edge computing: A fundamental computation-communication tradeoff study," \emph{IEEE Trans. Wireless Commun.}, vol. 19, no. 7, pp. 4563-4578, Jul. 2020. 
		
		\bibitem{TaoStudy}
		K.~{Li}, M.~{Tao}, and Z.~{Chen}, ``A computation-communication tradeoff study
		for mobile edge computing networks,'' in \emph{Proc. IEEE Int.
		Symp. Inf. Theory (ISIT)}, Paris, France, Jul. 2019, pp. 2639--2643.

		\bibitem{Osvaldo}
		J.~{Zhang} and O.~{Simeone}, ``On model coding for distributed inference and
		  transmission in mobile edge computing systems,'' \emph{IEEE Commun. Lett.}, vol.~23, no.~6, pp. 1065--1068, Jun. 2019.

		\bibitem{Kuikui}
		K.~{Li}, M.~{Tao}, J.~{Zhang}, and O.~{Simeone},  ``Multi-cell mobile edge coded computing: Trading communication and computing for distributed matrix multiplication," in \emph{Proc. IEEE Int. Symp. Inf. Theory (ISIT)}, Los Angeles, CA, Jun. 2020, pp.  215--220.

		\bibitem{SalimStair}
		R.~{Bitar}, P.~{Parag}, and S.~{El Rouayheb},  ``Minimizing latency for secure coded computing using secret sharing via staircase codes," \emph{IEEE Trans. Commun.}, vol. 68, no. 8, pp. 4609-4619, Aug. 2020.

		\bibitem{SalimRateless}
		R.~Bitar, Y.~Xing, Y.~Keshtkarjahromi, V.~Dasari, S.~El Rouayheb, and
		H.~Seferoglu,  ``PRAC: Private and rateless adaptive coded computation at the edge", in \emph{Proc. SPIE Defense + Commercial Sensing}, Baltimore, MD, May 2019.
		
				\bibitem{Dean}
		\BIBentryALTinterwordspacing
		J.~Dean and L.~A. Barroso, ``The tail at scale,'' \emph{Commun. ACM}, vol.~56,
		  no.~2, pp. 74--80, Feb. 2013.
		\BIBentrySTDinterwordspacing
		
		\bibitem{Mallick}
		\BIBentryALTinterwordspacing
		A.~Mallick, M.~Chaudhari, U.~Sheth, G.~Palanikumar, and G.~Joshi, ``Rateless
		  codes for near-perfect load balancing in distributed matrix-vector
		  multiplication,'' \emph{Proc. ACM Meas. Anal. Comput. Syst.}, vol.~3, no.~3, pp. 58:1--58:40, Dec. 2019.
		\BIBentrySTDinterwordspacing


				\bibitem{Zhang}
		J.~{Zhang} and O.~{Simeone}, ``Fundamental limits of cloud and cache-aided
		interference management with multi-antenna edge nodes,'' \emph{IEEE Trans. Inf. Theory}, vol.~65, no.~8, pp. 5197--5214, Aug.
		2019.

		\bibitem{Naderializadeh}
		N.~{Naderializadeh}, M.~A. {Maddah-Ali}, and A.~S. {Avestimehr},  ``Fundamental limits of cache-aided interference management," \emph{IEEE Trans. Inf. Theory}, vol. 63, no. 5, pp. 3092--3107, May 2017.
		


		
		\bibitem{Shamir}
		\BIBentryALTinterwordspacing
		A.~Shamir, ``How to share a secret,'' \emph{Commun. ACM}, vol.~22, no.~11, pp.
		  612--613, Nov. 1979.
		\BIBentrySTDinterwordspacing
		
	\end{thebibliography}
\end{document}